\documentclass[a4paper,preprintnumbers,notitlepage,superscriptaddress,amsmath,amssymb,nofootinbib]{revtex4-1}

\usepackage[utf8]{inputenc}
\usepackage[T1]{fontenc}
\usepackage{lmodern}
\usepackage[english]{babel}

\usepackage{color}
\usepackage{graphicx}
\usepackage{units}

\newcommand{\beq}{\begin{eqnarray}}
\newcommand{\eeq}{\end{eqnarray}}

\definecolor{coldw}{rgb}{0,0.6,0}
\definecolor{colfaded}{rgb}{0.6,0.6,0.6}

\newcommand{\Eout}{E_\mathrm{out}}

\begin{document}

\preprint{TTP12-047}
\preprint{SFB/CPP-12-100}


\title{A QCD description of the ATLAS jet veto measurement}

\author{Y. Hatta}\email{hatta@het.ph.tsukuba.ac.jp}
\affiliation{Faculty of Pure and Applied Sciences, University of Tsukuba, Tsukuba, Ibaraki 305-8571, Japan}
\author{C. Marquet}\email{cyrille.marquet@cern.ch}
\affiliation{Centre de physique th\'eorique, \'Ecole Polytechnique, CNRS, 91128 Palaiseau, France}
\author{C. Royon}\email{christophe.royon@cea.fr}
\affiliation{IRFU/Service de physique des particules, CEA/Saclay, 91191 Gif-sur-Yvette cedex, France}
\author{G. Soyez}\email{gregory.soyez@cea.fr}
\affiliation{Institut de physique th\'eorique, CEA/Saclay, 91191 Gif-sur-Yvette cedex, France}
\author{T. Ueda}\email{takahiro.ueda@kit.edu}
\affiliation{Institut f\"{u}r Theoretische Teilchenphysik, Karlsruhe Institute of Technology (KIT) D-76128 Karlsruhe, Germany}
\author{D. Werder}\email{dominik.werder@physics.uu.se}
\affiliation{Department of Physics and Astronomy, Uppsala University, Box 516, SE-751 20 Uppsala, Sweden}


\begin{abstract}
  We present a new QCD description of the ATLAS jet veto measurement,
  using the Banfi-Marchesini-Smye equation to constrain the inter-jet
  QCD radiation. This equation resums emissions of soft gluons at
  large angles, at leading-logarithmic accuracy, and accounts for both
  the so-called Sudakov and non-global logarithms. We show that this
  approach is able to reproduce, with no fitting parameters, the
  fraction of high-$p_T$ forward/backward di-jet events which do not
  contain additional hard emissions in the inter-jet rapidity
  range. We also compute the gap fraction in fixed-order perturbation
  theory to ${\mathcal O}(\alpha_s^2)$ and show that the perturbative
  series is unstable at large rapidity intervals.
\end{abstract}

\maketitle

\section{Introduction}

Recently the ATLAS collaboration measured, in proton-proton collisions
at the LHC, the fraction of di-jet events that do not contain additional
hard radiation in the inter-jet rapidity range \cite{Aad:2011jz}.
The original goal of this measurement was to look for BFKL-type effects,
as was previously done at the Tevatron with the so-called `jet-gap-jet'
observable \cite{d0,cdf,Chevallier:2009cu,Kepka:2010hu}. However the use
of a veto scale $\Eout\gg\Lambda_{QCD}$ by ATLAS, instead of a true rapidity
gap void of any hadronic activity, drastically reduces the sensitivity to
BFKL physics.
Rather, it turns out that this `jet-veto' measurement is sensitive to the
physics of inter-jet energy flow, and in the limit $p_T\gg \Eout$,
to the resummation of soft large-angle gluon emissions in perturbative QCD.

In that regime, the picture that emerges from the comparison
\cite{Aad:2011jz} between the ATLAS data and Monte-Carlo predictions
is far from clear: while HERWIG \cite{herwigpp} and PYTHIA \cite{pythia}
tend to be in reasonable agreement with the measurement, fixed-order
\cite{alpgen,powheg} calculations matched with parton shower are quite
below the data. Furthermore, adding BFKL logarithms, as done {\it
  e.g.} in HEJ \cite{hej}, to take into account the fact that the
rapidity difference between the two primary jets, $\Delta y$, can be
large does not help.
A possible origin of these shortcomings may be the insufficiency of
the angular-ordered parton shower to faithfully capture the physics of
energy flow. Indeed, the soft gluons which constitute inter-jet
radiation are not ordered in angle but rather in $p_T$, which makes
the relevant resummation single-logarithmic $(\alpha_s \ln
p_T/\Eout)^n$. 

The main goal of this paper is to provide a QCD-based computation of
the jet veto cross-section obtained by resumming these
logarithmically-enhanced $(\alpha_s \ln p_T/\Eout)^n$
terms\footnote{The fact that ATLAS quotes very small non-perturbative
  effects (smaller than 2\%) on their measurement of the gap fraction
  means that we may hope to achieve a good description of the jet veto
  cross-sections based purely on perturbative QCD.}.
There are actually two types of such logarithms~---~the Sudakov logs and the
non-global logs.
The former is more well-known and can be resummed by using the soft anomalous
dimension technique \cite{Oderda:1998en,Forshaw:2009fz}.
On the other hand, the non-global logs  arise  from
soft emissions from the secondary gluons
(not from the primary hard partons),
and can only be resummed in the large-$N_c$ limit \cite{Dasgupta:2001sh}  where the successive
emission of soft gluons may be viewed
as the splitting of color dipoles\footnote{This process is
  mathematically identical to the gluon splitting in the BFKL 
  evolution \cite{avsar}.}.
While HERWIG and PYTHIA can partly account for the effect of these logarithms
due to some overlap in the phase space \cite{Banfi:2006gy},
in principle they are not optimized for observables
such as the gap fraction.

In this paper we investigate whether the ATLAS data can be described
by a perturbative framework which incorporates the relevant
single-logarithms to all orders. 
For the resummation of the non-global logs, we follow the approach of
Banfi, Marchesini and Smye (BMS) \cite{Banfi:2002hw} who reduced the problem
to solving a nonlinear integro-differential equation.
In a previous paper \cite{Hatta:2009nd}, two of us numerically
solved this equation and estimated the survival probability of the BFKL-induced
rapidity gap.
Here we consider the one-gluon exchange
(octet) channel and compute the gap fraction in proton-proton collisions to be directly compared with the ATLAS data.
An approach similar in spirit was taken in \cite{DuranDelgado:2011tp,Forshaw:2009fz}
where the authors did not use the large-$N_c$ approximation and
fully included the Sudakov logs.
The non-global logs, however, were included only by introducing a $K$-factor.
For other recent discussions of energy flow and the ATLAS data,
see \cite{Deak:2011gj,Deak:2011ga,Alioli:2012tp,Gerwick:2012hq}.

The plan of the paper is as follows.
In Section II, we first compute the gap fraction in  fixed-order perturbation theory to ${\mathcal O}(\alpha_s^2)$  and point out a  problem which arises when the rapidity gap $\Delta y$ becomes large. In Section III, we describe the BMS approach which incorporates resummation and compute the gap fraction in this framework. The results are then compared with the ATLAS data.
Finally Section IV is devoted to conclusion.

\section{Fixed-order computation}

 The observable of interest is the gapped fraction in proton-proton collisions measured by the ATLAS collaboration at $\sqrt{s}=7$ TeV. It is   defined  by
\beq
{\mathcal R}(\Delta y,p_T)  \equiv  \frac{ d\sigma^\textnormal{veto} }{ d\Delta y\,d^2p_T }
	\Big/
 \frac{ d\sigma^\textnormal{incl} }{ d\Delta\, y d^2p_T }\,, \label{veto}
\eeq
where $\sigma^\textnormal{incl}$ is the inclusive cross section of di-jet events  and $\Delta y$ is the rapidity difference of the two jets which have mean transverse momentum  $p_T=(p_{T,1}+p_{T,2})/2 \ge 50$ GeV and rapidity $|y_i|<4.4$. Jets are reconstructed using 
the anti-$k_t$ algorithm \cite{antikt} with the radius parameter $R=0.6$.
In defining the di-jet system in each event,
ATLAS used two different selection criteria:
The highest-$p_T$ jet pair and the most forward/backward jet pair. 
 $\sigma^\textnormal{veto}$ is the gapped cross section in which a veto is applied to the di-jet cross section requiring that no jet with $p_T$ above $\Eout=20$ GeV is
observed in the rapidity interval between the two jets.

As explained in the introduction,
one expects in the perturbative computation of the ratio (\ref{veto})
large logarithms of the form $(\alpha_s \ln p_T/\Eout)^n$ to become important.
In particular, this includes the Sudakov-type logarithms $(\alpha_s \Delta y\ln p_T/\Eout)^n$ which are additionally enhanced by factors of $\Delta y$ when $\Delta y$ is large.  A fixed-order computation should then break down and the
large logarithms should be resummed in order to obtain a correct
description of  the vetoed di-jet cross section. 
However, before coming to that resummation, it remains interesting to
check around which value of  $\Delta y$ the
perturbative series becomes unstable. 
Therefore, in this section we first  perform a fixed-order study of the veto fraction. 

Since the leading-order (${\cal O}(\alpha_s^2)$) vetoed and inclusive
cross-sections are equal, the first non-trivial order for ${\mathcal R}$ is when
the cross-sections are computed at NLO.
At first sight, the fact that the inclusive jet cross-section is not
known at NNLO, because of its 2-loop pure-virtual contribution, seems
to indicate that the story ends here and ${\mathcal R}$ can only be computed
at ${\cal O}(\alpha_s)$. But, since this 2-loop pure-virtual
contribution to the jet cross-section is actually the same in the
vetoed and inclusive cases, one can show that, following the same arguments
as in \cite{xsect_ratio}, this unknown piece does not contribute to
the ${\cal O}(\alpha_s^2)$ term in the series expansion of ${\mathcal R}$. An
explicit series expansion shows that one can write ${\mathcal R}$ as\footnote{An
  alternative and maybe more direct approach is to write
  ${\mathcal R}=1-\sigma^\textnormal{noveto}/\sigma^\textnormal{incl}$ and to realize that taking both
  cross-sections at NLO, {\it i.e.}  $\sigma^\textnormal{noveto}$ at
  ${\cal O}(\alpha_s^4)$ and $\sigma^\textnormal{incl}$ at ${\cal O}(\alpha_s^3)$,
  would formally give a description of ${\mathcal R}$ at ${\cal O}(\alpha_s^2)$,
  while $\sigma^\textnormal{noveto}_{NLO}$ does not depend on the
  2-loop cross-section with only 2 partons in the final-state. The
  expression in \eqref{eq:ratio_series} is no more than an explicit
  series expansion of that statement.}
\begin{equation}\label{eq:ratio_series}
{\mathcal R}(\Delta y, p_T)
 = 1
 - \frac{\left(\frac{d^2\sigma^\textnormal{noveto}}{dp_T\,d\Delta y}\right)_{LO}}
        {\left(\frac{d^2\sigma^\textnormal{incl}}{dp_T\,d\Delta y}\right)_{LO}} \alpha_s
 + \left[
   \frac{\left(\frac{d^2\sigma^\textnormal{noveto}}{dp_T\,d\Delta y}\right)_{LO}
         \left(\frac{d^2\sigma^\textnormal{incl}}{dp_T\,d\Delta y}\right)_{NLO}}
        {\left(\frac{d^2\sigma^\textnormal{incl}}{dp_T\,d\Delta y}\right)_{LO}^2}
 - \frac{\left(\frac{d^2\sigma^\textnormal{noveto}}{dp_T\,d\Delta y}\right)_{NLO}}
        {\left(\frac{d^2\sigma^\textnormal{incl}}{dp_T\,d\Delta y}\right)_{LO}}
   \right] \alpha_s^2
\end{equation}
where the $\alpha_s$ dependence has been explicitly factored out and
$\sigma^\textnormal{noveto}=\sigma^\textnormal{incl}-\sigma^\textnormal{veto}$.

\begin{figure}
\includegraphics[width=\textwidth]{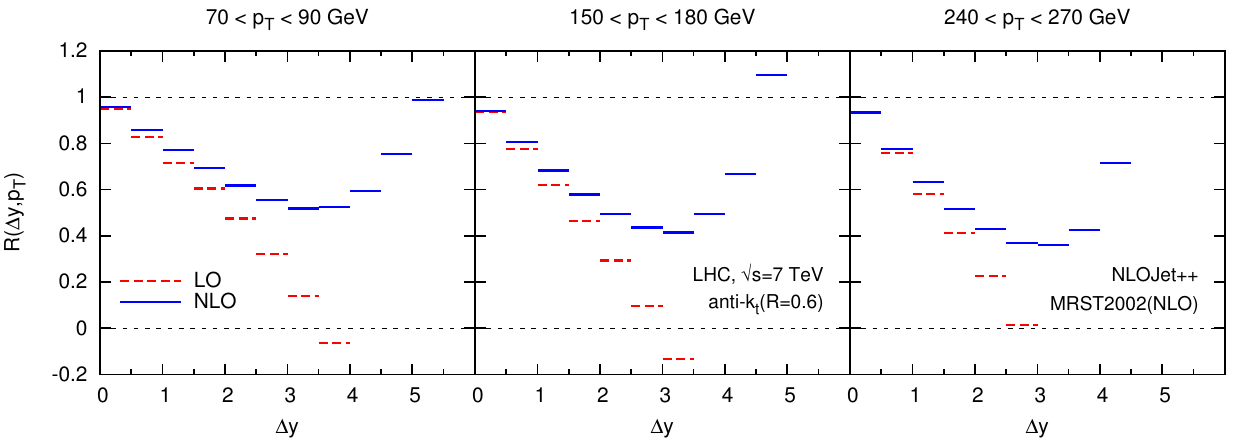}
\caption{LO (${\cal{O}}(\alpha_s)$) and NLO ${\cal O}(\alpha_s^2)$
  predictions for the fraction of di-jet events with a jet
  veto.}\label{fig:fixed_order}
\end{figure}

All the differential cross-sections in \eqref{eq:ratio_series} can be
computed explicitly using NLOJet++ \cite{nlojet}. 
Following the ATLAS analysis \cite{Aad:2011jz} as described above and focusing, for definiteness, on the case
where the di-jet system is made of the most forward and most backward
jets, we simulated events with NLOJet++ v4.1.2  using
the anti-$k_t$ algorithm as implemented in FastJet
\cite{fastjet}. In the perturbative computation we have set both the
renormalization and factorization scales\footnote{Although this would
  naturally apply for a di-jet system based on the two hardest jets in
  the event, $p_T$ is the natural scale choice also for this
  forward-backward configuration which is expected to be dominated by
  a $t$-channel exchange. Note also that we could have studied the
  scale uncertainty and compared it to the differences between the LO
  and NLO curves in Fig.~\ref{fig:fixed_order}.}  to $p_T$ and we have
used the MRST2002 PDF set at NLO~\cite{mrst2002}.

The result of this analysis is shown on Fig.~\ref{fig:fixed_order}
where ${\mathcal R}$ is plotted as a function of the rapidity interval $\Delta y$
for different bins in $p_T$. The dashed (red) curves are obtained by
truncating \eqref{eq:ratio_series} at order $\alpha_s$ while the solid
(blue) curves also keep the ${\cal O}(\alpha_s^2)$ terms. We immediately notice that the LO result becomes negative when
$\Delta y$ becomes large while the NLO result becomes larger than
1. These unphysical results are actually not surprising and stem from the fact that at large
rapidity interval $\Delta y$ one expects a Sudakov-type behavior of the form
${\mathcal R}\propto \exp(-C\alpha_s\,\Delta y \ln p_T)$ which, truncated at fixed order,
would give ${\mathcal R}\propto 1-C\alpha_s\,\Delta y \ln p_T + (C\alpha_s\,\Delta
y \ln p_T)^2/2+\cdots$.
More interestingly, one sees explicitly from
Fig.~\ref{fig:fixed_order} that the perturbative series becomes
completely unstable for $\Delta y\gtrsim 2$, with a weak dependence on 
$p_T$. The main message of this study is that one should thus expect
some form of resummation to be necessary for $\Delta y \gtrsim 2$ which is practically the whole interesting region of the physics of rapidity gap.

\section{Resummed computation}

In this section we undertake the resummed computation of the gapped fraction ${\mathcal R}$ which should cure the problem with the fixed-order computation we have just seen. Our calculation is based on the approach by Banfi-Marchesini-Smye (BMS). We first describe their approach and adapt it for the problem at hand, and then present the numerical results.

\subsection{The BMS equation}
\label{sec:bms}

\begin{figure}
\includegraphics[width=0.4\linewidth]{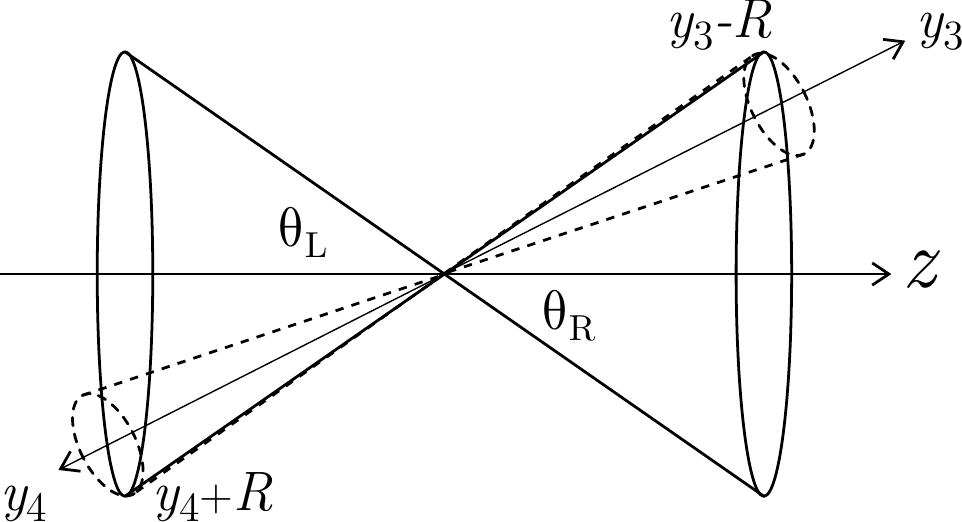}
\caption{
$\theta_{R/L}$ ($y_{R/L}=\ln \cot \theta_{R/L}/2$) is the opening angle
(rapidity)  of the right/left cone.
}
\label{fig:f1}
\end{figure}

Consider a pair of cones pointing back-to-back in the direction of the beam
axis ($z$-axis) as in Fig.~\ref{fig:f1}.
The opening angles of the right cone $\theta_R$ and the left cone $\theta_L$
need not be the same.
A quark and an antiquark in the color singlet state (`dipole') are contained
in the cones and moving in the directions $\Omega=(\theta,\phi)$ and
$\Omega'=(\theta',\phi')$, respectively, with the transverse momentum $p_T$.
Let $P_\tau(\theta_R,\theta_L,\Omega,\Omega')$ be the probability that the
total amount of energy, or the transverse momentum, emitted from the dipole
into the region outside the cones is less than $\Eout$.
 The `evolution parameter' $\tau$ is related to $\Eout$. Taking into account the running of the coupling,
%
%
we have\footnote{
With a fixed coupling approximation, we find $\tau = (\alpha_s\,N_c/\pi)\,\ln(p_T/\Eout)$.
} \cite{Dasgupta:2001sh,Marchesini:2003nh}:
\beq
\tau
  = \int_{\Eout}^{p_T}\frac{dk_T}{k_T} \frac{\alpha_s(k_T)N_c}{\pi}
  = \frac{1}{2 b} \ln \left(\frac{\alpha_s(\Eout)}{\alpha_s(p_T)}\right)\,,
\label{eq:alphas-run}
\eeq 
where $\alpha_s(p_T) = \pi/( 2b\,N_c\ln(p_T/\Lambda_{QCD}))$, and
$b=(11N_c-2n_f)/12N_c$ with $n_f=5$ being the number of active
flavors.

%
To zeroth order, $P=1$. Within the large-$N_c$ approximation, BMS derived a nonlinear integro-differential equation in $\tau$ \cite{Banfi:2002hw} which resums all the single logarithms (both Sudakov and non-global) 
$\left(\alpha_s \ln p_T/\Eout\right)^n \sim \tau^n$ which appear in the weak coupling expansion of $P$. 
The numerical solution to this equation has been previously obtained in
\cite{Banfi:2002hw,Hatta:2009nd}.
Here we solve it again for the kinematic range relevant to the ATLAS data.
In doing so, we switch to the rapidity variable
\beq
y= \ln \cot \frac{\theta}{2}\,,
\eeq
instead of the polar angle $\theta$ in the arguments of $P$.
Also, we suppress the dependence on the azimuthal angle $\phi$
because $P$ is a function of the difference  $\Delta \phi=\phi-\phi'$ and
below we need solutions only at $\Delta\phi=\pi$.
Since the probability $P$ must not depend on a specific Lorentz frame,
we can evaluate it in the di-jet c.m.s.
\beq
P_\tau(y_R,y_L,y,y')
= P_{\tau}\left(
	\frac{y_R-y_L}{2},
	-\frac{y_R-y_L}{2},
	y-\frac{y_R+y_L}{2},
	y'-\frac{y_R+y_L}{2}
\right)\,. \label{boost}
\eeq
Eq.~(\ref{boost}) allows us to restrict ourselves to the symmetric configuration
of the cones,
which will greatly facilitate the computations below.

\subsection{The jet-veto cross section in hadron-hadron collisions}

As already suggested by BMS, one can utilize the above probability $P_\tau$ to  constrain the inter-jet radiation in hadron-hadron collisions. The initial and final state soft radiations  from a given partonic subprocesses may be deconstructed, in the large-$N_c$ approximation, into those from elementary dipoles by inspecting the flow of color. Consider the simplest $p_1p_2\to p_3p_4$ process:  $qq' \to qq'$ or
$ \bar{q}\bar{q}'\to \bar{q}\bar{q}'$ (different quark flavors).
The partonic cross section is given by 
\beq
\frac{d\sigma_{qq'}}{d\hat{t}} = \frac{1}{16\pi \hat{s}^2}h^A(\hat{s},\hat{t},\hat{u})
\eeq
where $\hat{s}$, $\hat{t}$, and $\hat{u}$ are the standard partonic Mandelstam variables  ($\hat{s}=x_1x_2s$, etc) and ($C_F=(N_c^2-1)/2N_c=\frac{4}{3}$)
\beq
&&h^A(s,t,u)=g^4\frac{C_F}{N_c}\left(\frac{s^2+u^2}{t^2}\right)\,.
\eeq 
At large-$N_c$,  and in the one-gluon exchange, color flows as $1\to 4$ and $2\to 3$. Thus the radiation
from the four-parton system $(1234)$ factorizes into that from two dipoles
$(14)$ and $(23)$.
We require that the amount of energy emitted in the central region
bounded by the edges of the jets $p_3$ and $p_4$ with jet radius $R$
is less than $\Eout$ (see, Fig.~\ref{fig:f1}).
The cross section with this requirement is given by \cite{Banfi:2002hw}
\beq
\frac{d\sigma^\textnormal{veto}_{qq'}}{d\hat{t}}
	= \frac{1}{16\pi \hat{s}^2} \Bigl(h^A(\hat{s},\hat{t},\hat{u})
	P_\tau(y_3-R,y_4+R,\infty,y_4)P_\tau(y_3-R,y_4+R,y_3,-\infty) \nonumber \\
+ h^A(\hat{s},\hat{u},\hat{t})
	P_\tau(y_3-R,y_4+R,\infty,y_3)P_\tau(y_3-R,y_4+R,y_4,-\infty) \Bigr)\,.
\label{qq-01}
\eeq
Note that we have added the `$u$-channel' term although there is no
$u$-channel diagram for the process $qq' \to qq'$, because,
when calculating the gapped cross section, we must sum over the two
cases $\Delta y=y_3-y_4>0$ and $\Delta y <0$ since $q$ and $q'$ are not identical
particles.
Adding these two contributions effectively amounts to adding the
$u$-channel term but restricting to the $\Delta y>0$ case,
facilitating later computation.

By boost invariance (\ref{boost}), we have the relations
\beq
&& P_\tau(y_3-R,y_4+R,\infty,y_4)
	= P_\tau\left(\frac{\Delta y}{2}-R, -\frac{\Delta y}{2}+R,
	\infty,-\frac{\Delta y}{2}\right) \equiv P_{14}(\tau,\Delta y)\,, \nonumber \\
&&P_\tau (y_3-R,y_4+R,y_3,-\infty)
	= P_\tau\left(\frac{\Delta y}{2}-R,-\frac{\Delta y}{2}+R,
	\frac{\Delta y}{2},-\infty\right)
\equiv P_{23}(\tau,\Delta y)\,,
\eeq
\beq
&& P_\tau(y_3-R,y_4+R,\infty,y_3)
	= P_\tau\left(\frac{\Delta y}{2}-R,-\frac{\Delta y}{2}+R,
	\infty, \frac{\Delta y}{2}\right) \equiv P_{13}(\tau,\Delta y)\,, \nonumber \\
&&  P_\tau(y_3-R,y_4+R,y_4,-\infty)
	= P_\tau\left(\frac{\Delta y}{2}-R,-\frac{\Delta y}{2}+R,
	-\frac{\Delta y}{2}, -\infty\right) \equiv P_{24}(\tau,\Delta y)\,.
\eeq
Clearly, $P_{14}=P_{23}$ and $P_{13}=P_{24}$, so that (\ref{qq-01}) takes the simpler form
\beq
\frac{d\sigma^\textnormal{veto}_{qq'}}{d\hat{t}}
	&=&\frac{1}{16\pi \hat{s}^2} \Bigl(h^A(\hat{s},\hat{t},\hat{u})
	P_{14}P_{23}
	+ h^A(\hat{s},\hat{u},\hat{t})  P_{13}P_{24} \Bigr)
	\nonumber \\
&=& \frac{1}{16\pi \hat{s}^2} \Bigl(h^A(\hat{s},\hat{t},\hat{u})
	P_{14}^2
	+ h^A(\hat{s},\hat{u},\hat{t})  P_{13}^2 \Bigr)\,.
\label{qq-02}
\eeq
In Appendix we list the gapped cross section of all the other partonic subprocesses obtained in a similar way. The results are convoluted with the PDFs to give the hadronic gapped cross section 
\beq
\frac{d\sigma^\textnormal{veto}}{d\Delta y \, d^2p_T }
	= \sum_{ij}^{q,\bar{q},g} \int_{Y_{\min}}^{Y_{\max}} dY \,
        x_1f_i(x_1, p_T) \, x_2f_j(x_2, p_T)\,
	\frac{1}{\pi} \frac{d\sigma^\textnormal{veto}_{ij}}{d\hat{t}}\,,
	\label{1}
\eeq
 where $Y=\frac{y_3+y_4}{2}$ and $Y_{\max}=-Y_{\min}=4.4$ for the ATLAS setup.

\subsection{Results}

\begin{figure}
\includegraphics[scale=0.75]{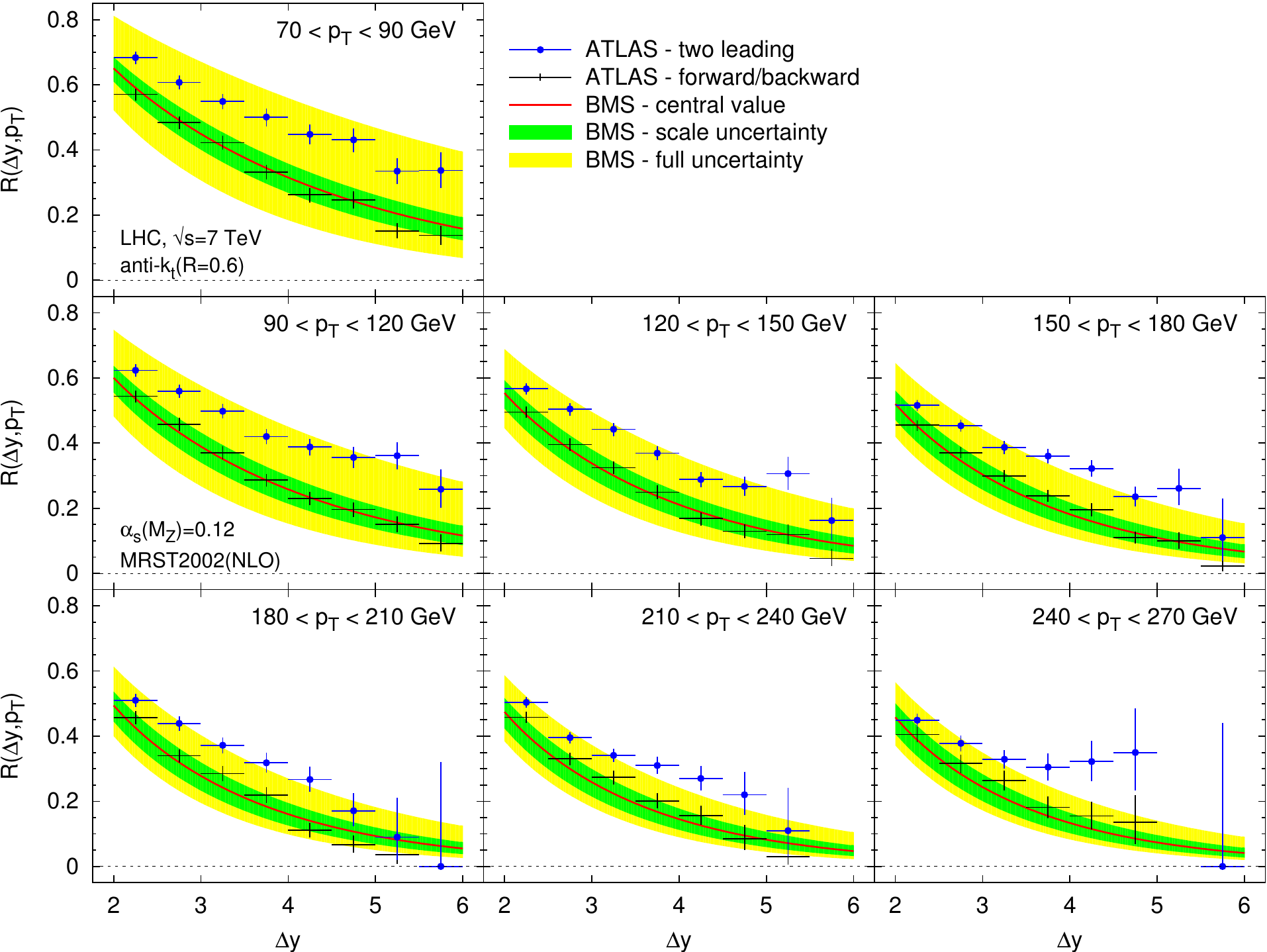}
\caption{Comparison of the resummed veto fraction with the ATLAS
  measurement, for a fixed veto energy of $\Eout=\unit[20]{GeV}$, in
  different bins of $p_T$. The inner (green) uncertainty band is
  obtained taking into account only the renormalization and
  factorization scale uncertainties, while the outer (yellow) band
  also includes the subleading logarithmic uncertainty. For the ATLAS
  data, circles represent the case where the two leading jets are
  selected while the one where the most forward and backward jets are
  selected are represented by crosses.}
\label{fig:set-best-2}
\end{figure}

\begin{figure}
\includegraphics[scale=0.75]{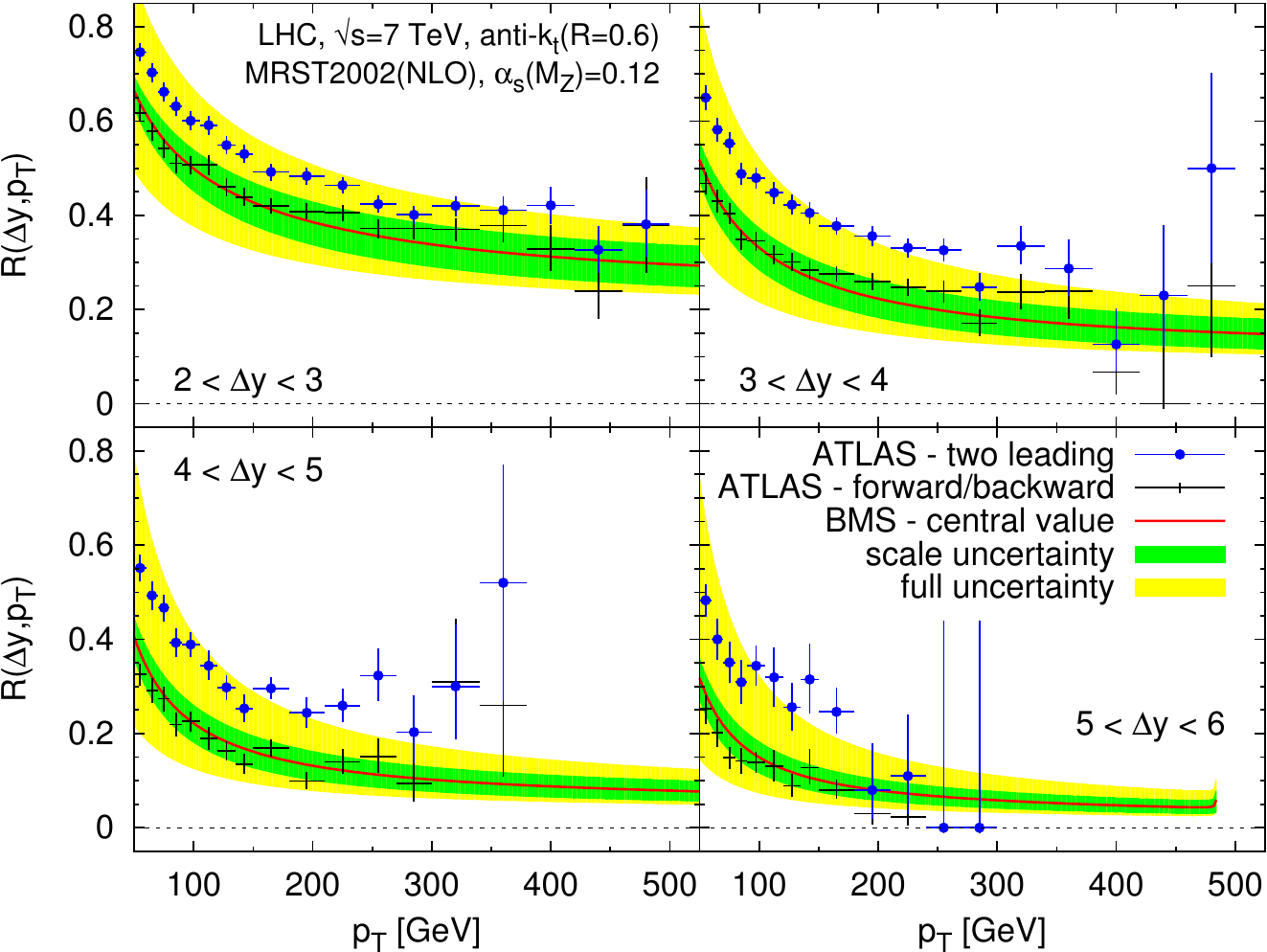}
\caption{Comparison of the resummed veto fraction with the ATLAS
  measurement, for a fixed veto energy of $\Eout=\unit[20]{GeV}$, in
  different bins of $p_T$. See Fig. \ref {fig:set-best-2} for
  details.}
\label{fig:set-best-1}
\end{figure}

We compute the resummed gapped fraction ${\mathcal R}$ by dividing (\ref{1}) by the leading-order  cross section (that is,  the same expression except that $P$'s are set to 1 everywhere). 
%
%
We have kept $n_f=5$ as adequate in the kinematic range under
consideration and fixed the running of the coupling by imposing
$\alpha_s(M_Z)=0.12$.
The theoretical uncertainty on our predictions comes from different
sources: First, the renormalization and factorization scale
uncertainties are obtained by varying the scale of $\alpha_s$ in the
matrix elements and in the definition of $\tau$
(Eq. \eqref{eq:alphas-run}) --- in the latter case, we vary directly
the scale of $\alpha_s(k_T)$ in the integrand --- as well as the scale
$p_T$ in the PDFs by a factor of 2 up and down. We exclude the cases
where the renormalization scale $\mu_r$ and the factorization scale
$\mu_f$ are $(\mu_r,\mu_f) = (p_T/2, 2p_T)$ or $(2p_T, p_T/2)$. It is
interesting to note that the factor $\alpha_s^2(\mu_r)$ appearing in
the matrix elements cancels when taking the cross-section ratio and
thus the only renormalization scale uncertainty that remains
is the one in the definition of $\tau$.
Then, the uncertainty related to subleading logarithmic corrections
that are not resummed by the leading $\alpha_s^n\,\ln^n(p_T/\Eout)$
series is estimated by varying the upper integration bound in the
definition of $\tau$, Eq. \eqref{eq:alphas-run} by a factor of 2 up
and down.
Finally, the uncertainty band is then taken as the envelope of all the
resulting curves.

The results are compared against the ATLAS measurements as published
in \cite{Aad:2011jz}.  Since we use the LO cross section where there
are only two jets in the final state, we cannot distinguish the two
ATLAS data sets based on different di-jet selection criteria. While
naively we expect our predictions to be in better agreement with the
``two-leading jets'' selection, we shall see below that we better
reproduce the data based on the most forward/backward jets.

Fig.~\ref{fig:set-best-2} shows the dependence on the jet rapidity
separation for different bins in jet $p_T$.  We note that the model is
overall in good agreement with the measurement. Needless to say, the
unphysical behavior, ${\mathcal R}>1$ or ${\mathcal R}<0$, of the
fixed-order computation (see Fig.~\ref{fig:fixed_order}) has
disappeared. In Fig.~\ref{fig:set-best-1}, the result is presented
over the jet $p_T$ for different bins in $\Delta y$.
%
In both cases, the agreement with the ATLAS data is good, especially
with the sample obtained by selecting the most forward/backward
jets. The predictions systematically undershoot the data based on the
two leading jets although they remain in agreement once the
uncertainty due to subleading effects are taking into account.

\begin{figure}
\includegraphics[scale=0.75]{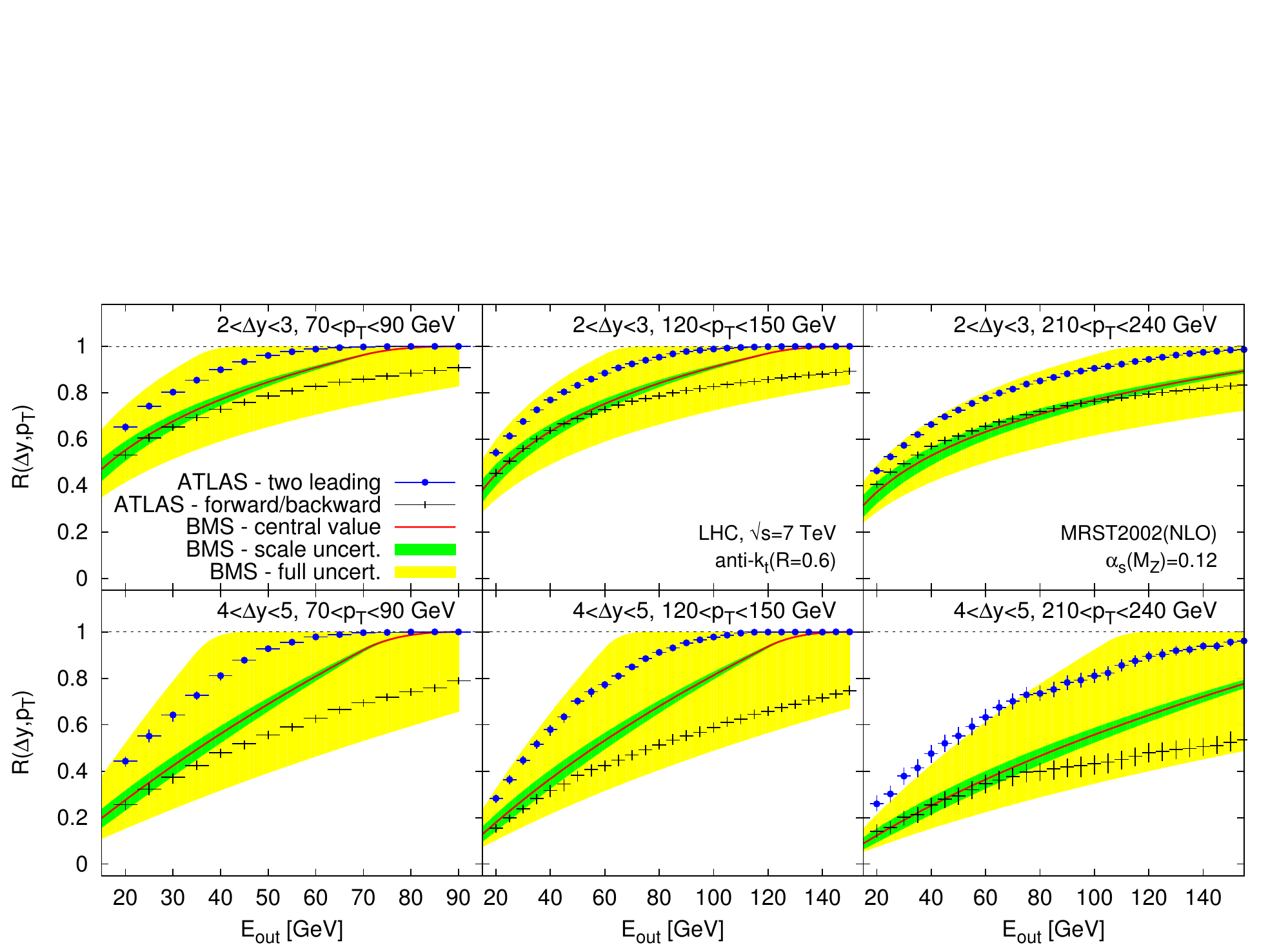}
\caption{Comparison of the resummed veto fraction with the ATLAS
  measurement, for different kinematic bins, as a function of the veto
  threshold $\Eout$. See Fig. \ref {fig:set-best-2} for details.}
\label{fig:set-best-3}
\end{figure}

Then, Fig.~\ref{fig:set-best-3} demonstrates the dependence on the jet
veto threshold energy $\Eout$.  We note first that in our results the
veto fraction saturates to unity as the threshold $\Eout$ approaches
the jet $p_T$ scale, as expected. In the data, it falls short of unity
because of the NLO ($2\to 3$) corrections.  On the other hand, the
agreement is best for the smaller values of $\Eout$, which is expected
as well since the formalism we use requires a large scale separation
between the jet $p_T$ and the threshold $\Eout$.


Finally, it is worth mentioning that uncertainties in the choice of
the PDF set mostly cancel in the ratio (\ref{veto}). We checked this
by trying two different sets, MRST2002~\cite{mrst2002} and CT10
\cite{Lai:2010vv}, and found no noticeable difference.

\section{Conclusions and Outlook}

In this paper, we have investigated the QCD resummation of the Sudakov
and the non-global logarithms induced by soft gluon emissions in the
context of the jet veto cross-section.
We conclude that the ATLAS measurement is well described by tree-level
QCD supplemented with the jet veto probability calculated
perturbatively using the BMS equation which resums both logarithms
mentioned above.
Actually, the impact of the non-global logarithms is modest compared
to that of the Sudakov logarithms: with $\Eout$ as large as 20 GeV,
we estimate that the non-global contribution reduces the gap fraction
by about 15\%.

Having said this, we must comment on uncertainties in our results
other than the scale uncertainties already examined.
Above all, our description does not allow to disentangle between the
two methods proposed by ATLAS to identify the di-jet system. While our
approach better reproduces the data where the most forward and the
most backward jets are selected, it is quite below the measurement
obtained by selecting the two hardest jets in the event. This somewhat
goes against the naive expectation that our description should better
reproduce the latter situation.
%
This probably means that effects other than the ones included in the
BMS equation are at play.
Still, we have shown that the resummation of the soft gluon
emissions gives a reasonable description of the jet veto probability.

Various additional effects would play a role if we wanted to improve
our predictions. 
First, our approach resums the soft gluon emissions at the
leading logarithmic accuracy, so subleading effects would
potentially be important. [A rough estimate of this is shown by the yellow band in the figures.] 
Then, we do not include $1/N_c$ corrections neither in the Sudakov
logarithms nor in the non-global ones. The $1/N_c$ corrections in the
Sudakov logarithms are fully taken into account in
\cite{DuranDelgado:2011tp}, but those for the non-global ones pose a
serious theoretical challenge.
Also, the NLO ($2 \to 3$) corrections to the hard parton cross
sections can be important, especially at small $\Delta y$ and for
discriminating the two ATLAS data sets corresponding to two different
definitions of the di-jet system. 
In principle, our resummed approach can be extended to $2 \to 3$
processes, but the flow of color will be considerably more complicated
and it is not clear to us at the moment if such an extension is
practical. Alternatively, it may be more practical to match the BMS
predictions to the fixed-order predictions from Section II which are
expected to provide more accurate NLO corrections at least in the
small $\Delta y$ region.
Finally, we have seen that the data are well described by the color octet
contribution alone, without introducing additional BFKL-like (singlet)
contributions.
As already mentioned in the introduction, the ATLAS choice $\Eout\gg
\Lambda_{QCD}$ significantly reduces the sensitivity to BFKL
exchanges. Presumably, we need to go to lower values of $\Eout$
\cite{Hatta:2009nd}, thus approaching the case of a perfect gap
($\Eout \to 0$) \cite{Chevallier:2009cu,Kepka:2010hu,Marquet:2012ra},
to enter the BFKL-dominated regime. It is nevertheless interesting to
mention that these corrections would tend to increase our results for
the gap fraction ${\cal {R}}$, especially at large $\Delta y$: Our
estimates for ${\cal {R}}$ based on a pure-BFKL calculation are in the
range 0.8-0.9, and the HEJ event generator, which includes BFKL
effects, also predicts values larger than the ones obtained by
ATLAS. A combination of the BMS and BFKL contributions could thus
bring our predictions in better agreement with the ATLAS measurement
obtained based on the two hardest jets in the event. We leave this for
future study.

\begin{acknowledgments}
We thank G. Ingelman, R. Peschanski and G. Salam for discussions and comments.
The work of T.U. was supported by the DFG through SFB/TR 9
``Computational Particle Physics''.
The calculations were partially performed on the HP XC3000
at Steinbuch Centre for Computing of KIT.
This work was partially supported by the French Agence Nationale de la
Recherche, under the grant ANR-10-CEXC-009-01.
\end{acknowledgments}

\appendix

\section{Parton subprocess}
In this Appendix we list the gapped cross section of all the partonic subprocesses which contribute to the sum in (\ref{1}). Some of them were already considered in \cite{Banfi:2002hw}. The others are simply obtained by crossing symmetry.
 We shall need the following building blocks of the cross section 
 \cite{Marchesini:1987cf,Banfi:2002hw}
 \beq
&&h^B(s,t,u)=g^4\frac{C_F}{N_c}
	\left(\frac{s^2+u^2}{t^2} + \frac{2}{N_c}\frac{s}{t}\right)\,,
	\label{hb}\\
&&h^C(s,t,u)=g^4C_F \frac{u}{t}\left(\frac{t^2+u^2}{s^2}-\frac{1}{N_c^2}\right)\,,\\
&&h^D(s,t,u)=2g^4 \frac{N_c^2}{N_c^2-1}
	\left(1-\frac{tu}{s^2} -\frac{su}{t^2} + \frac{u^2}{st}\right)\,,
\eeq
 and the solutions to the BMS equation for different dipole configurations 
\beq
&&P_\tau(y_3-R,y_4+R,y_3,y_4) = P_\tau\left(\frac{\Delta y}{2}-R,-\frac{\Delta y}{2}+R,\frac{\Delta y}{2},-\frac{\Delta y}{2}\right) \equiv P_{34} (\tau,\Delta y)\,,
\\
&& P_\tau(y_3-R,y_4+R,\infty,-\infty) = P_\tau \left(\frac{\Delta y}{2}-R,-\frac{\Delta y}{2}+R,\infty,-\infty\right) \equiv P_{12}(\tau,\Delta y)\,.
\eeq 
Note that $P_{34}$ is evaluated at $\Delta \phi = \phi_3-\phi_4=\pi$ as appropriate for back-to-back jets. \\

\begin{itemize}

\item $q\bar{q}' \to q\bar{q}'$ (different quark flavors) \\
This can be obtained from $qq'\to qq'$ by crossing $s\leftrightarrow u$ and $2\leftrightarrow 4$
\beq
\frac{d\sigma^\textnormal{veto}_{q\bar{q}'}}{d\hat{t}} &=&\frac{1}{16\pi \hat{s}^2} \Bigl(h^A(\hat{u},\hat{t},\hat{s})
P_{12}P_{34}
+ h^A(\hat{t},\hat{u},\hat{s})  P_{12}P_{34} \Bigr)\,,
\eeq
where we added a $u$--channel diagram (by the same reason as in (\ref{qq-01}))
which is obtained by crossing $t\leftrightarrow u$, $3\leftrightarrow 4$.

\item
$q\bar{q} \to q'\bar{q}'$ (different quark flavors)\\
Obtained from $qq' \to qq'$ via crossing $s\leftrightarrow t$, $2\leftrightarrow 3$
\beq
\frac{d\sigma^\textnormal{veto}_{q\bar{q}\to q'\bar{q}'}}{d\hat{t}} &=&\frac{1}{16\pi \hat{s}^2} \Bigl(h^A(\hat{t},\hat{s},\hat{u})
P_{14}P_{23}
+ h^A(\hat{u},\hat{s},\hat{t})  P_{13}P_{24} \Bigr)
\nonumber \\
&=&\frac{1}{16\pi \hat{s}^2} \Bigl(h^A(\hat{t},\hat{s},\hat{u})
P_{14}^2
+ h^A(\hat{u},\hat{s},\hat{t})  P_{13}^2 \Bigr)\,.
\eeq

\item $qq \to qq$, $\bar{q}\bar{q}\to \bar{q}\bar{q}$ (the same flavor) \cite{Banfi:2002hw}: \\
\beq
\frac{d\sigma^\textnormal{veto}_{qq}}{d\hat{t}} &=& \frac{1}{16\pi \hat{s}^2}
\Bigl(h^B(\hat{s},\hat{t},\hat{u})
P_{14}P_{23}
+ h^B(\hat{s},\hat{u},\hat{t})  P_{13}P_{24} \Bigr)
\nonumber \\
&=& \frac{1}{16\pi \hat{s}^2}
\Bigl(h^B(\hat{s},\hat{t},\hat{u})
P^2_{14}
+ h^B(\hat{s},\hat{u},\hat{t})  P_{13}^2 \Bigr)\,.
\eeq
Compare with (\ref{qq-02}).
Since the final state particles are identical, one integrates over only half
of the phase space, that is, only the $\Delta y>0$ case.
Note that the interference term between the two diagrams for $qq\to qq$ is subleading in $N_c$ and does
not correspond to a process with definite color flow.
It was distributed symmetrically into the second term of
$h^B$ (\ref{hb}) \cite{Marchesini:1987cf}.
Similar comments apply to the other subleading contributions in $h^C$ and $h^D$ below.

\item $q\bar{q} \to q\bar{q}$: \\
 Obtained from $qq\to qq$ by crossing $s\leftrightarrow u$, $2\leftrightarrow 4$ and adding the $\Delta y<0$ contribution ($t\leftrightarrow u$, $3\leftrightarrow 4$) because of nonidentical particles in the final state
\beq
\frac{d\sigma^\textnormal{veto}_{q\bar{q}\to q\bar{q}}}{d\hat{t}} &=& \frac{1}{16\pi \hat{s}^2}\Bigl(h^B(\hat{u},\hat{t},\hat{s})P_{12}P_{34} + h^B(\hat{u},\hat{s},\hat{t})P_{13}P_{24} + h^B(\hat{t},\hat{u},\hat{s})P_{12}P_{34} + h^B(\hat{t},\hat{s},\hat{u})P_{14}P_{23}\Bigr) \nonumber \\
&=& \frac{1}{16\pi \hat{s}^2}\Bigl(h^B(\hat{u},\hat{t},\hat{s})P_{12}P_{34} + h^B(\hat{u},\hat{s},\hat{t})P_{13}^2+h^B(\hat{t},\hat{u},\hat{s})P_{12}P_{34} + h^B(\hat{t},\hat{s},\hat{u})P_{14}^2\Bigr)\,.
\eeq

\item  $q\bar{q} \to gg$ \cite{Banfi:2002hw}:\\
\beq
\frac{d\sigma^\textnormal{veto}_{q\bar{q}}}{d\hat{t}}&=& \frac{1}{16\pi \hat{s}^2}\Bigl( h^C(\hat{s},\hat{t},\hat{u})P_{34} P_{13}P_{24}+h^C(\hat{s},\hat{u},\hat{t})P_{34} P_{14}P_{23} \Bigr)
\nonumber \\
&=& \frac{1}{16\pi \hat{s}^2}\Bigl( h^C(\hat{s},\hat{t},\hat{u})P_{34} P_{13}^2+h^C(\hat{s},\hat{u},\hat{t})P_{34} P_{14}^2 \Bigr)\,.
\eeq

\item $gg\to q\bar{q}$: \\
The same as $q\bar{q} \to gg$ except for the color factor $(3/8)^2$ and the addition of the $\Delta y<0$ contribution ($t\leftrightarrow u$, $3\leftrightarrow 4$)
\beq
\frac{d\sigma^\textnormal{veto}_{gg\to q\bar{q}}}{d\hat{t}} = \frac{1}{16\pi \hat{s}^2} \left(\frac{3}{8}\right)^2
\Bigl(h^C(\hat{s},\hat{t},\hat{u})P_{34} P_{13}^2 +h^C(\hat{s},\hat{u},\hat{t})P_{34} P_{14}^2 \Bigr)\times 2\,.
\eeq

\item $qg \to qg$ and $\bar{q}g \to \bar{q}g$: \\
 Obtained from $q\bar{q}\to gg$ by crossing $s\leftrightarrow t$, $2\leftrightarrow 3$, multiplying the color factor $3/8$, adding the $\Delta y<0$ contribution ($t\leftrightarrow u$, $3\leftrightarrow 4$)
\beq
\frac{d\sigma^\textnormal{veto}_{qg}}{d\hat{t}} &=& \frac{-1}{16\pi \hat{s}^2}\frac{3}{8} \Bigl( h^C(\hat{t},\hat{s},\hat{u})P_{24}P_{12}P_{34} + h^C(\hat{t},\hat{u},\hat{s} ) P_{24}P_{14}P_{23}+ h^C(\hat{u},\hat{s},\hat{t})P_{23}P_{12}P_{34} + h^C(\hat{u},\hat{t},\hat{s} ) P_{23}P_{13}P_{24}\Bigr) \nonumber \\
&=& \frac{-1}{16\pi \hat{s}^2}\frac{3}{8} \Bigl( h^C(\hat{t},\hat{s},\hat{u})P_{13}P_{12}P_{34} + h^C(\hat{t},\hat{u},\hat{s} ) P_{13}P_{14}^2+ h^C(\hat{u},\hat{s},\hat{t})P_{14}P_{12}P_{34} + h^C(\hat{u},\hat{t},\hat{s} ) P_{14}P_{13}^2\Bigr)\,.
\eeq
Note the overall minus sign.

\item $gg \to gg$  \cite{Banfi:2002hw}: \\
\beq
\frac{d\sigma^\textnormal{veto}_{gg \to gg}}{d\hat{t}} &=& \frac{1}{16\pi \hat{s}^2} \Bigl(
h^D(\hat{s},\hat{t},\hat{u})P_{12}P_{13}P_{24}P_{34} + h^D(\hat{s},\hat{u},\hat{t})P_{12}P_{14}P_{23}P_{34} +h^D(\hat{u},\hat{t},\hat{s}) P_{14}P_{24}P_{13}P_{23}\Bigr)
\nonumber \\
&=& \frac{1}{16\pi \hat{s}^2} \Bigl(
h^D(\hat{s},\hat{t},\hat{u})P_{12}P_{13}^2P_{34} + h^D(\hat{s},\hat{u},\hat{t})P_{12}P_{14}^2P_{34} +h^D(\hat{u},\hat{t},\hat{s}) P_{14}^2P_{13}^2\Bigr)\,.
\eeq

\end{itemize}

\end{document}